\def\GRL {{Geophys.\ Res.\ Lett.~}}

\def\PhP  {{Phys. Plasmas~}}
\def\PRL {{Phys. Rev. Lett.~}}

\def\PRep  {{Phys. Rep.~}}
\def\PR  {{Phys. Rev.~}}
\def\PRE  {{Phys. Rev. E~}}
\def\PSr {{Phys. Scr.~}}
\def\s {{Science~}}
\def\NPG {{Nonlinear Processes in Geophys.~}}

\def\erf {\mathop{\rm erf}\nolimits}
\def\erfi {\mathop{\rm erfi}\nolimits}

\def\rmax {r_{\rm max}}
\documentclass[aps,prl,showpacs,twocolumn,groupedaddress,superscriptaddress]{revtex4}
\usepackage{graphicx}

\def\vr {{\bf r}}

\begin{document}

%\title{Three-dimensional Bernstein-Greene-Kruskal Electron Solitary Waves
%in magnetized plasma}
\title{Width-amplitude relation of Bernstein-Greene-Kruskal solitary
  waves}

\author{Li-Jen Chen}

\affiliation{Department of Physics and Astronomy, University of Iowa,
  Iowa City, IA 52242-1479}

\author{David J. Thouless}

\affiliation{Department of Physics, University of Washington, Seattle,
  WA 98195-1560}

\author{Jian-Ming Tang}

\affiliation{Department of Physics and Astronomy, University of Iowa,
  Iowa City, IA 52242-1479}

\date{\today}

\begin{abstract}
  Inequality width-amplitude relations for three-dimensional
  Bernstein-Greene-Kruskal solitary waves are derived for magnetized
  plasmas.  Criteria for neglecting effects of nonzero cyclotron radius
  are obtained.  We emphasize that the form of the solitary potential
  is not tightly constrained, and the amplitude and widths of the
  potential are constrained by inequalities.  The existence of a
  continuous range of allowed sizes and shapes for these waves makes
  them easily accessible. We propose that these solitary waves can be
  spontaneously generated in turbulence or thermal fluctuations. We
  expect that the high
  excitation probability of these waves should alter the bulk
  properties of the plasma medium such as electrical resistivity and
  thermal conductivity.
\end{abstract}


\maketitle

%MAIN TEXT
Coherent structures with nonuniform charge densities
are ubiquitous in plasma systems.
Laboratory experiments have shown that such structures can be
generated by applying voltage pulses \cite{Lynov79,Bachet01}, voltage
jumps \cite{Chan84}, intense laser \cite{Montgomery01} or plasma beam
injections \cite{IAI00}.  Increasing numbers of space-borne
observations have revealed frequent appearance of electrostatic solitary
structures in space plasmas (\cite{Ergun98,Jolene03} and references
therein) including regions where magnetic reconnection occurs
\cite{Drake03,Matsumoto03}.  Solitary waves can efficiently transport
energy, momentum and charge, and are one of the building blocks in
a deterministic description of turbulence \cite{Saffman95,IBM80}. The
study of their allowed parameter space is crucial in establishing
their relevance to real systems.  Most solitary waves, such as those
for shallow water (Korteweg-de Vries solitons) and those that describe
crystal dislocations (sine-Gordon solitons), have a strict one-to-one
mapping between their widths, amplitudes and velocities
\cite{Drazin83}. However, in collisionless plasmas, the widths and
amplitudes of the electrostatic solitary waves that exhibit vortex
structures in phase space \cite{Schamel86} are not tightly constrained
\cite{Chen-npg,Chen-grl}. Although the width-amplitude relation of
these so-called Bernstein-Greene-Kruskal (BGK) electron solitary waves
in three dimensions (3D) have been studied by a few authors
\cite{Muschietti02,Chen-npg,Chen-grl}, the results were limited either by
imcomplete analysis of the width-amplitude relation
\cite{Muschietti02} or by incomplete
solutions \cite{Chen-npg,Chen-grl}, and the broad impact has not been
recognized.  In this letter, we derive inequality
width-amplitude relation for 3D BGK solitary waves. The description is
valid for both electron and ion modes.  The inequality width-amplitude
relation dictates a continuous range of allowed sizes and amplitudes
for these waves, and thus makes them easily accessible.  We propose
that the continua of allowed existence range enable BGK solitary waves
to be spontaneously generated in thermal fluctuations as well as
turbulence, and is responsible for
their ubiquitous presence in widely different classes of collisionless
plasmas.

To construct exact nonlinear solutions that are localized in 3D, we
use the BGK approach that was formulated for 1D nonlinear
Vlasov-Poisson equations \cite{BGK57}, but extend the Poisson equation
to 3D. We construct azimuthally symmetric solutions in the limit of
infinite magnetic field, and then tune down the magnetic field to
obtain the criteria for neglecting effects of nonzero cyclotron radius.
One key step in the BGK approach is to separate particles that are
trapped in the potential and those that are passing.  We prescribe the
potential form and the passing particle distribution, solve for the
trapped particle distribution, and derive the physical parameter
range.  This approach is much easier than to prescribe the passing and
trapped particle distributions to solve for the potential
\cite{Schamel82,Schamel86}, and so allows us to explore the solution space much
more fully.

We consider two species of charge carriers (electrons and one type of
ions), each with charge $q_s$, mass $m_s$, and thermal energy $T_s$.
The background magnetic field ${\bf B}$ is along the $\hat z$
direction.  In the strong field limit, particles move only along ${\bf
  B}$ with velocity $v$, and the distribution functions $f_s$ satisfy
the following Vlasov equations,
\begin{eqnarray}
v\frac{\partial f_s(\vr,v)}{\partial z}-\frac{q_s}{m_s}\frac{\partial \Phi(\vr)}
{\partial z}\frac{\partial f_s(\vr,v)}{\partial v} & = & 0 \;,
\label{FF}
\end{eqnarray}
where $\Phi(\vr)$ is the electrostatic potential. It can be easily shown
by chain rules that any $f(\vr_\perp,w)$ is a solution to Eq.~(\ref{FF})
since its dependence on $z$ and $v$ is only through the particle
energy $w=m_sv^2/2+q_s\Phi(\vr)$. Such distribution functions and the
potential are further constrained by the Poisson equation,
\begin{eqnarray}
-\nabla^2\Phi(\vr) & = & \sum_{s=1}^2 \int_{q_s\Phi(\vr)}^\infty dw\frac{4\pi n_sq_sf_s(\vr_\perp,w)}{\sqrt{2m_s[w-q_s\Phi(\vr)]}} \;,
\label{PP}
\end{eqnarray}
where the velocity space integral has been converted to energy space
integral, and $f_s(\vr_\perp,w)$ is normalized so that $n_s$ is the
particle density in the unperturbed region where charge neutrality
gives $\sum_sn_sq_s=0$.  Species $1$ is defined to be the one which
involves trapping ($\min\, [q_1\Phi(\vr)] <0$), and species $2$ does
not. The distribution function $f_1$ is further divided into passing
and trapped components, $f_p$ and $f_{tr}$. The second term in
Eq.~(\ref{FF}) is nonlinear as $\Phi$ is a functional of the particle
distributions and vice versa.  In physical terms, the system is
nonlinear because
plasma particles collectively determine the
mean-field potential, and the potential in turn determines how
particles distribute themselves.  It is the presence of this nonlinear
term that admits solitary wave solutions which exhibit localized
structures in potentials and distribution functions.

Eq.~(\ref{FF}) can be thought as a set of 1D Vlasov equations in the
$\hat z$ direction for given $\vr_\perp$.  These parallel Vlasov equations
are coupled by the perpendicular profile of the potential $\Phi$ through
Eq.~(\ref{PP}). If $\Phi$ is known, Eq.~(\ref{PP}) reduces to a set of 1D
integral equations parameterized by $\vr_\perp$. For given $f_2$ and the
passing distribution $f_p$, the trapped distribution $f_{tr}$ can be
found by solving these integral equations. The important requirement
for the solutions to be physical is that the trapped distribution
$f_{tr}$ so determined should be nonnegative.  This leads to a
self-consistent constraint on the form of the potential specified at
the beginning.  It turns out that neither the potential forms nor the
passing distributions are tightly constrained.  One can prescribe
different localized potential functions or different passing particle
distributions (as long as the distribution functions satisfy the
Vlasov equation). As an illustrating example, the solitary potential
is chosen to be an azimuthally symmetric double Gaussian,
\begin{eqnarray}
\Phi(r,z) & = & g\psi\,\exp\left(-z^2/2\delta_z^2-r^2/2\delta_r^2\right) \;,
\label{phi}
\end{eqnarray}
where $g=-{\rm sign}(q_1)$ in order for $\Phi$ to trap species 1 particles,
 $\psi$ is the potential amplitude and is
positive, $r=|\vr_\perp|$, $\delta_z$ and $\delta_r$ are the parallel and
perpendicular widths.  The distributions $f_p$ and $f_2$ are of the
Boltzmann type,
\begin{eqnarray}
f_p(w) & = & \sqrt{2m_1/\pi T_1}\exp(-w/T_1) \;,\\
f_2(w) & = & \sqrt{2m_2/\pi T_2}\exp(-w/T_2) \;.
\label{fp}
\end{eqnarray}
By carrying out the integrals of $f_p$ and $f_2$ in Eq.~(\ref{PP}), we
obtain the trapped particle density,
\begin{widetext}
\begin{eqnarray}
n_{tr}(\Phi) & = & \Phi\left[\frac{r^2}{\delta_r^2}\left(\frac{1}{\delta_r^2}-\frac{1}{\delta_z^2}\right)
-\frac{2}{\delta_r^2}-\frac{1}{\delta_z^2}-\frac{2}{\delta_z^2}\ln\left(\frac{\Phi}{g\psi}\right)\right]
-e^{-\Phi}\left[1-\erf(\sqrt{-\Phi})\right]+\exp{(-t\Phi)},
\label{ntr}
\end{eqnarray}
where $t=q_2 T_1/(q_1 T_2)$. To simplify the expression, we have set
the length unit to be $\lambda_D = \sqrt{T_1/ 4\pi n_1 q_1^2}$, energy unit
$T_1$ and charge unit $q_1$ ($\Phi$ becomes strictly negative
in units of $T_1/q_1$). We have also rewritten the expression in terms of the
potential, a crucial step for analytically solving the integral
equation. We refer readers to Appendix B in Ref.~\cite{Lijen02} for
the details of solving this Volterra type integral equation, and simply write
down the result here,
\begin{eqnarray}
\frac{f_{tr}(r,w)}{\sqrt{2m_1}} & = & \frac{2\sqrt{-w}}{\pi}\left[\frac{r^2}{\delta_r^2} \left(\frac{1}{\delta_r^2}-\frac{1}{\delta_z^2}\right) - \frac{2}{\delta_r^2} + \frac{1}{\delta_z^2} -\frac{2}{\delta_z^2}\ln\frac{4w}{-\psi} \right] + \frac{e^{-w}}{\sqrt{\pi}} \left[1-\erf(\sqrt{-w}) \right] -\frac{e^{tw}}{\sqrt{\pi}}\sqrt{t}\erfi(\sqrt{-tw}) \;,
\label{ftr}
\end{eqnarray}
\end{widetext}
where $w<0$, and $\erfi(z)=\erf(i z)/i$ is the complex error function
which is a real function of its argument.

At this juncture, it is instructive to consider the small amplitude
behavior of our solution and compare with previous results. 
Expand the RHS of
Eq. (\ref{ntr}) for small $\Phi$.  The leading term comes from the
passing particle density and is of order $\sqrt{\Phi}$, and the next
order is linear in $\Phi$.  This means that for small potential
amplitudes, the nonlinearity actually dominates over the linear
response. In this case, even when $T_i\sim T_e$ for electron-proton
plasma, the trapped distribution
$f_{tr}$ can be solved keeping only the leading term, and it yields a
constant $\sqrt{2m_1/\pi}$, which demonstrates that the plasma can
sustain the solitary structure by providing a uniform trapped particle
distribution.  This result, in the context of ion holes, sharply
contrasts the previous result \cite{Schamel86} which predicts that ion
holes do not exist when $T_e/T_i < 3.5$.

In recapitulation, we have made one step forward to obtain solutions
in Eq.~(\ref{ftr}) for arbitrary $t$ taking into account density
perturbations of both species. We also obtained new understanding:
nonlinearity dominates over linear response in the small amplitude
limit -- an unexpected result as one normally expects nonlinearity to
be unimportant for small amplitudes.  Hereafter, we shall use electron
holes as the example to discuss the solution behavior and demonstrate
how an inequality width-amplitude relation is obtained in the $t\to 0$
limit, the limit which all previous analytical solutions for electron
holes were based on. This limit is also applicable to ion holes generated
in laboratory experiments where $T_i/T_e\sim 0.03$ to 0.1 \cite{Chan84,IAI00}.

\begin{figure}
  \includegraphics[width=8.5cm]{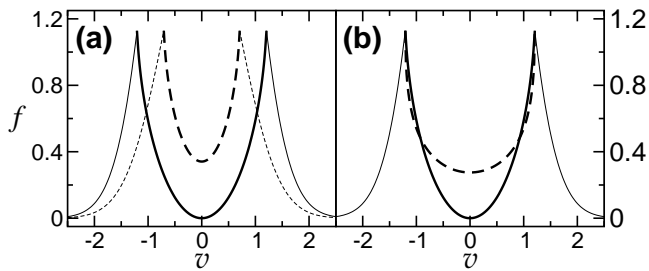}
\caption{Velocity distributions of particles at the center ($r=0, z=0$) of
  BGK solitary structures for different potential amplitudes and
  sizes demonstrating how the tuning of these parameters affects
  particle distributions. Parameters are: (a) $(\psi, \delta_r, \delta_z)=(1.45, 5, 3)$ (solid lines) and $(\psi, \delta_r,
  \delta_z)=(0.5, 5, 3)$ (dashed lines), (b) $(\psi, \delta_r, \delta_z)=(1.45,5,3)$
  (solid lines) and $(\psi, \delta_r, \delta_z)=(1.45, 5,10)$ (dashed lines). In
  both cases the thick lines represent trapped particle distributions,
 and the thin lines are passing particle distributions. }
\label{fv}
\end{figure}
We illustrate below how the tuning of the parameters ($\psi,\delta_r,\delta_z$)
affects the trapped particle distribution.  Fig.~\ref{fv} plots
$f_{tr}$ (thick lines) and $f_p$ (thin lines) at $r=0$ and $z=0$ as a
function of velocity $v$. The solid curves in both (a) and (b)
correspond to a BGK solitary wave that has zero phase space density at
its phase space center ($r=0,z=0,v=0$).  When the size of the
structure is fixed, decreasing the amplitude raises the center phase
space density as shown by the dashed curve in Fig.~\ref{fv}(a). On the
other hand, increasing the amplitude would lower the center phase
space density from zero to a negative value (not shown), and hence
result in unphysical solutions. When the amplitude is fixed,
increasing the parallel size raises the center phase space density as
shown in Fig.~\ref{fv}(b) by the dashed curve. Varying $\delta_r$ results
in a similar effect.

We now proceed to derive the width-amplitude inequality relation.  For
$\delta_r \leq \delta_z$, the first term in Eq.~(\ref{ftr}) is positive, hence the
global minimum is $f_{tr}(r=0,w=-\psi)$ and its being nonnegative
determines the corresponding width-amplitude relation.  For $\delta_r >
\delta_z$, the global minimum of $f_{tr}$ occurs at maximum allowed $r$.
For a given $w<0$, the maximum $r$ at which a trapped particle with
energy $w$ can exist is the $\rmax$ that satisfies $-w=\Phi(\rmax,0)$.
Putting Eq.~(\ref{phi}) into this condition, we obtain $\rmax^2 =
-2\delta_r^2 \ln(-w/\psi)$.  Since the condition $f_{tr}(\rmax,w)\geq 0$
guarantees $f_{tr}(r\leq \rmax,w)\geq 0$ for $\delta_r > \delta_z$, we replace $r^2$
in $f_{tr}(r,w)$ by $\rmax^2$.  The global minimum of
$f_{tr}(\rmax,w)$ is $f_{tr}(\rmax,w=-\psi)$, hence the condition
$f_{tr}(\rmax,w=-\psi)\geq 0$ yields the width-amplitude relation for $\delta_r >
\delta_z$.
\begin{figure}
\includegraphics[width=4.5cm]{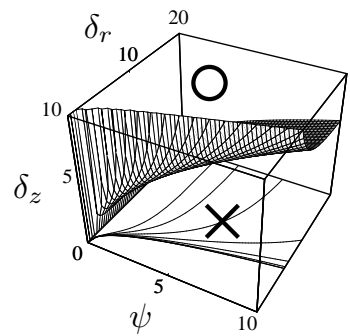}
\caption{The inequality relation between the parallel size ($\delta_z$),
  the perpendicular size ($\delta_r$) and the potential amplitude ($\psi$)
  showing that there is a continuum of allowed heights and widths for
  BGK solitary waves. The allowed region, the shaded surface and
  above, is marked by {\Large $\circ$} and the forbidden by $\times$.  The
  curves on the $\delta_z=0$ plane are projections of constant $\delta_z$
  contours on the surface.}
\label{wa}
\end{figure}

The conditions $f_{tr}(r=0,w=-\psi) \geq 0$ when $\delta_r \leq \delta_z$ and
$f_{tr}(\rmax,w=-\psi)\geq 0$ when $\delta_r > \delta_z$ yield exactly the same
expression.  Upon re-arrangement, the resulting inequality is written
as
\begin{eqnarray}
\delta_z & \geq & \sqrt{{2(4\ln 2 -1)\over{{\sqrt{\pi}e^\psi[1-\erf(\sqrt{\psi})]/\sqrt{\psi}}-{4/\delta_r^2}}}} \;.
\label{dzineq}
\end{eqnarray}
Fig.~\ref{wa} plots this inequality.  Parameters lying on or above the
shaded surface are allowed ({\Large $\circ$}), and those under the surface
are forbidden ($\times$).  Curves on the $\delta_z=0$ plane are projections of
the constant $\delta_z$ contours to help visualization of the trend of the
shaded surface.  The shaded surface curves up toward infinity because
the denominator in inequality~(\ref{dzineq}) has to be positive, and
that yields another relation between $\delta_r$ and $\psi$ (given by $\delta_r^2 >
4\sqrt{\psi}/(\sqrt{\pi} e^\psi[1-\erf\sqrt{\psi}])$ and is plotted as the
asymptotic curve on $\delta_z=0$ plane).  These inequalities occur because
we are free to place the global minimum of $f_{tr}$ in a continuous
range by correspondingly adjusting the amplitude and widths.  A point
on the shaded surface represents a parameter set that yields zero
phase space density at $w=-\psi$, that is $(r=0,z=0,v=0)$, the center of
the solitary phase space structure.  One example of the empty-centered
distribution has been provided in Fig.~\ref{fv}.  Lowering the
amplitude or increasing the widths shifts a point on the surface to
the region above, and Figs.~\ref{fv}(a) and (b) illustrate
respectively the effects on the distribution functions.

We note that in the limit of $\delta_r \to \infty$, inequality (\ref{dzineq})
reduces to the width-amplitude relation for 1D BGK solitary waves.
This limit of inequality (\ref{dzineq}) gives an upper bound for $\delta_z$
that is valid for all finite $\delta_r$.
The resulting 1D inequality relation provides us a ground to
understand the discrepancy of whether the width should increase
\cite{Turikov84} or decrease \cite{Schamel82} with the amplitude.
Both results are contained in the inequality relation with that by
Ref.~\cite{Turikov84} corresponding to the lower bounding curve since only
empty-centered distributions were studied, and that by
Ref.~\cite{Schamel82} contained in the region above the curve as the
distributions only take finite values in the center of the phase space
structure.

To establish the validity of the above results in finite magnetic
field, we need to know whether the effect of nonzero cyclotron radius
would result in decoherence of the solitary structure, that is, how
the distance of the instantaneous particle guiding center to the symmetry axis would
vary.  Therefore, it is best to look at the particle motion projected
onto the 2D $x-y$ plane perpendicular to the magnetic field.  The
motion of a charged particle inside the solitary structure is
influenced by the uniform ${\bf B}$ and the 3D inhomogeneous ${\bf E}$
which upon projection becomes time-dependent ${\bf E}_{2D}$. It can be
shown that when the time variation scale of ${\bf E}_{2D}$ is much
smaller than the cyclotron frequency $\omega_c$, and the spatial variation
scale ($L$) of ${\bf E}_{2D}$ is much larger than the cyclotron radius
$r_c$, the instantaneous guiding center would spiral around the
infinite-field guiding center, and the solitary structure can be
maintained (details of calculations will be presented elsewhere
\cite{PhP}).  Since the time variation of ${\bf E}_{2D}$ is
characterized by the frequency ($\omega_b$) of the parallel
trapped-particle bouncing in the potential, the condition can be
written as
\begin{eqnarray}
\omega_b/\omega_c \ll 1 & \longrightarrow & \sqrt{m_e\psi/e}/(B \delta_z)\ll 1 \\
r_c/L \ll 1 & \longrightarrow & \sqrt{2m_e\psi/e}/(B\delta_r) \ll 1 \;,
\end{eqnarray}
where we have expressed on the right hand side of the arrows the
condition in terms of familiar variables. 
For relevant numerical investigation of nonzero cyclotron radius effects, 
the readers are referred to Ref.~\cite{Muschietti02}.

We note that the size and the amplitude of BGK solitary waves do not
have a lower cut-off within our theory.  The underlying reason is that
the screening of the charged core is accomplished by trapped particles
which are part of the solitary structure itself. Debye screening is not
involved in these self-consistent, self-sustained nonlinear objects.
Their size can be well below the Debye radius as far as there are
enough particles in the solitary wave to ensure the validity of the
mean-field approach. Taking a Debye radius ($\lambda_D$) 100 m and a plasma
density 5 cm$^{-3}$ (typical of the low altitude auroral ionosphere),
a width of 0.01 $\lambda_D$ for the solitary potential allows $5\times 10^6$
particles in the structure. Indeed, sub-Debye scale solitary waves
have been observed \cite{Ergun98}.

It is the primary purpose of this paper to address the physical origin
and significance of the inequality width-amplitude relation plotted in
Fig.~\ref{wa}.  The freedom to continuously adjust the global minimum
of the trapped particle distribution is due to the collisionless
nature of the plasma.  Collisionlessness preserves the identity of
trapped and passing particles as the energy of a particle is
conserved. Collisions destroy energy conservation of individual
particles, and consequently, do not allow the existence of trapped
particle state nor the adjustment on the occupation number of the
state. Therefore, the kinetic solitary waves have a continuum of
allowed potential heights and widths, in great distinction to fluid
solitons.  Moreover, even with the same ambient plasma distribution,
different functional forms for the solitary potential is allowed.  The
impact of this multitude of continua of allowed potentials is that
these BGK states can be easily excited. In a system with certain
fluctuation level and different fluctuation lengths, BGK states may be
accessed easily since for a fixed amplitude there is a wide range of
allowed widths. We therefore propose that in turbulent systems, BGK
solitary waves can be spontaneously generated in the absence of
two-stream or current-driven instabilities.  The spontaneous
generation of these coherent structures and their subsequent mutual
interaction may dominate the transport properties of the turbulence.
Because these solitary waves have many degrees of freedom, energy,
momentum and charge are readily transferred between them,
so that they can make important contributions to bulk properties of the
plasma such as thermal transport and electrical resistivity. At the
vicinity of the waves, passing particles are accelerated and, together
with trapped particles, form counterstreaming beams (Fig.~1). Hence
the velocity spread increases significantly and results in much higher
average velocity spread (heating)
of the plasma.  Moreover, particle trapping
will prevent particles from free acceleration by the
applied electric field and
regulate the electric current. The high excitation probability of BGK
waves can thus lead to finite resistivity that is required for melting the
frozen-in magnetic flux and facilitate reconnection to occur in
collisionless plasmas.

In summary, we have derived inequality width-amplitude relation for 3D
BGK solitary waves, and established their relevance in finite magnetic
fields.  The continuum of allowed potential sizes and shapes of the
waves is due to their kinetic nature, and leads to their ubiquitous
presence.  We envision that BGK solitary waves can be spontaneously
generated in thermal fluctuations as well as in turbulence.

The research at the University of Iowa is supported in part
by the DOE Cooperative
Agreement No. DE-FC02-01ER54651 and NSF ATM 03-27450,
and at the University of Washington by NSF DMR-0201948.

\end{document}